# Characterization of phases and boundary effects in U(1) gauge theory *

Werner Kerler[a], Claudio Rebbi[b], and Andreas Weber[a]

[a]Fachbereich Physik, Universität Marburg,
D-35032 Marburg, Germany

[b]Department of Physics, Boston University,
Boston, MA 02215, USA

We show that the two phases of the 4-dimensional compact U(1) lattice gauge theory are characterized by the existence or absence of an infinite current network, defining "infinite" on a finite lattice in a manner appropriate to the chosen boundary conditions. In addition for open and fixed boundary conditions we demonstrate the effects of inhomogeneities and provide examples of the reappearance of an energy gap.

## 1. INTRODUCTION

The widely accepted first order nature of the phase transition of the 4-dimensional compact U(1) lattice gauge theory has recently been questioned by the authors of Ref. [1] who implemented the theory on the surface of a 5-dimensional cube rather than on the usual 4-dimensional torus. The suggestion there was that on a manifold with trivial homotopy group the energy gap disappears. This has initiated other investigations which instead of periodic boundary conditions have used fixed boundary conditions [2], suppression of monopoles at the boundaries [3] and open boundary conditions [4]. In the present contribution we investigate the effects of periodic, open and fixed boundary conditions in more detail.

## 2. CHARACTERIZATION OF PHASES

Our results show that the two phases of the theory are characterized by the existence or absence of an infinite current network, defining "infinite" on finite lattices in accordance with the chosen boundary conditions. We find that this characterization holds independently of the boundary conditions and provides an order parameter which is superior to the one [5] based on the relative size of the largest network $n_{\text{max}}/n_{\text{tot}}$.

In the case of periodic boundary conditions we define "infinite" by "topologically nontrivial in all directions". While for individual loops the topological characterization is straightforward, for the networks of loops a more sophisticated analysis [8, 4] becomes necessary. Figure 1 gives the probability $P_{\text{net}}$ for the occurrence of a network nontrivial in all directions as a function of $\beta$. It is obvious that this is an order parameter which takes the values 0 and 1 for cold and hot phase, respectively, and which compares favorably with $n_{\text{max}}/n_{\text{tot}}$.

We have checked that referring simply to the extension of the largest network is not adequate for periodic boundary conditions. The meaning of "infinite" in this case would be that for each direction the projection of the network (to one dimension) should cover the full lattice extension. In our test on the $16^4$ lattice, where the peaks of the energy distribution are well separated, in the cold phase we have obtained the value 0.069(16) for this order parameter, as opposed to 0.000 for $P_{\text{net}}$ considered above. Thus it is neither size nor extension but topology which is relevant in the case of periodic boundary conditions.

*Contribution to LATTICE 95, International Symposium on Lattice Field Theory, Melbourne, Australia, 1995. Supported in part under DFG grants Ke 250/7-2 and 250/12-1 and under DOE grant DE-FG02-91ER40676.

For open boundary conditions the lattice is no longer self-dual. On the dual lattice the current lines (i.e. the lines with nonzero currents) may end at the boundaries. Now "infinite" is defined by "touching the boundaries in all directions".

Figure 1 shows that the related probability $P_{\rm net}$ provides an order parameter which again takes the values 0 and 1 for the phases, respectively, and compares favorably with $n_{\rm max}/n_{\rm tot}$.

Fixed boundary conditions are obtained by starting from periodic ones and putting the gauge group elements $U = 1$ at the boundary. The surface of the lattice then is made up of 3-dimensional cubes with all elements put to one. On the dual lattice these correspond to links with vanishing monopole currents. The dual set accessible to current lines is thus just a lattice with open boundary conditions (while the original lattice with fixed boundary conditions is homeomorphic to the sphere $\mathbf{S}^4$). Now "infinite" is defined by "reaching the boundaries in all directions". As can be seen from Figure 1 the related probability $P_{\rm net}$ again provides an order parameter which takes the values 0 and 1 for the phases, respectively, and compares favorably with $n_{\rm max}/n_{\rm tot}$.

It is also obvious from Figure 1 that for open and fixed boundary conditions the transition regions are much larger than for periodic boundary conditions. This may be explained by considerably larger finite-size effects present in these cases.

## 3. EFFECTS OF INHOMOGENEITIES

In our simulations we have determined the plaquette energy distributions $P(E)$ as well as the monopole density distributions. In the following we show the results for $P(E)$ at the phase transition. The corresponding monopole density results are very similar. Figure 2 shows $P(E)$ obtained on the $16^4$ lattice. For periodic boundary conditions it exhibits a gap with well separated peaks while for open and fixed boundary conditions there is no gap.

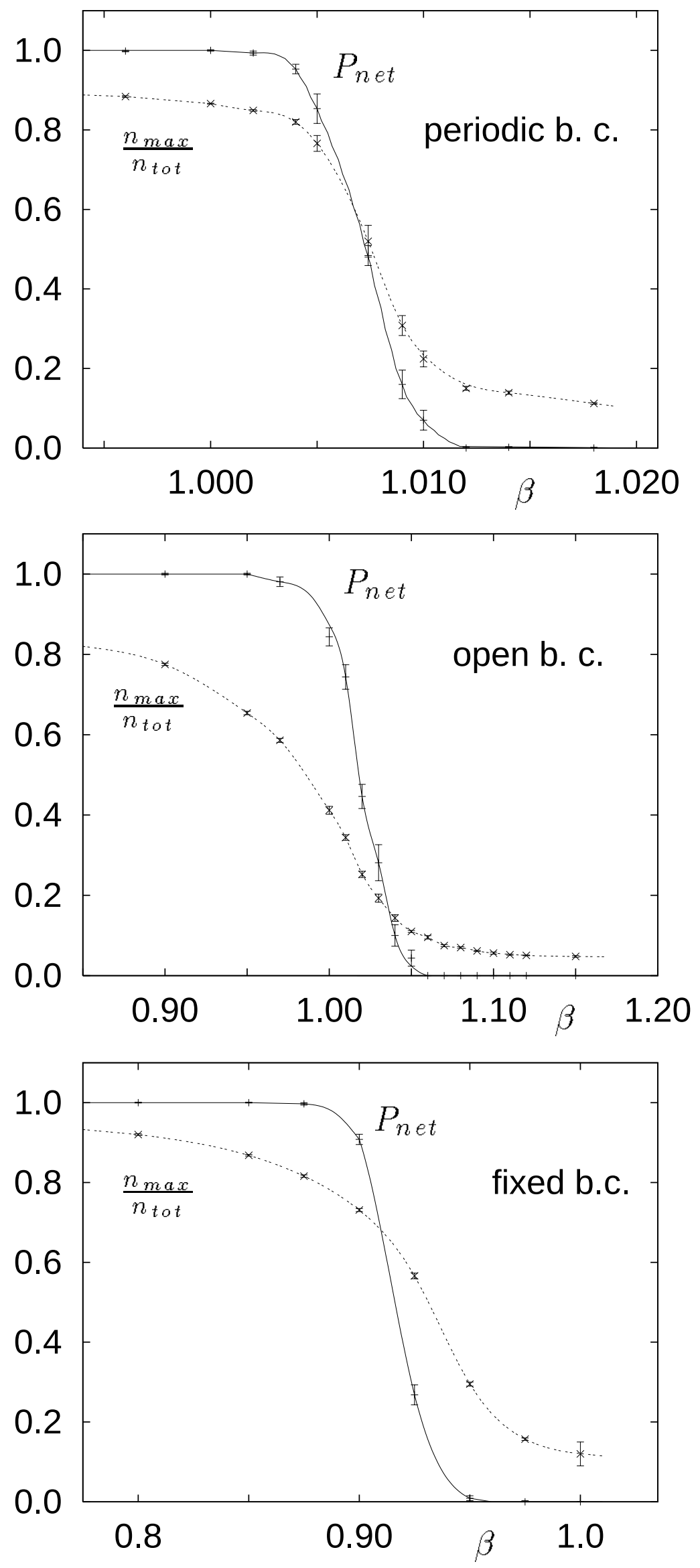

Figure 1. Order parameters $P_{net}$ and $n_{max}/n_{tot}$ as functions of $\beta$ on $8^4$ lattice.

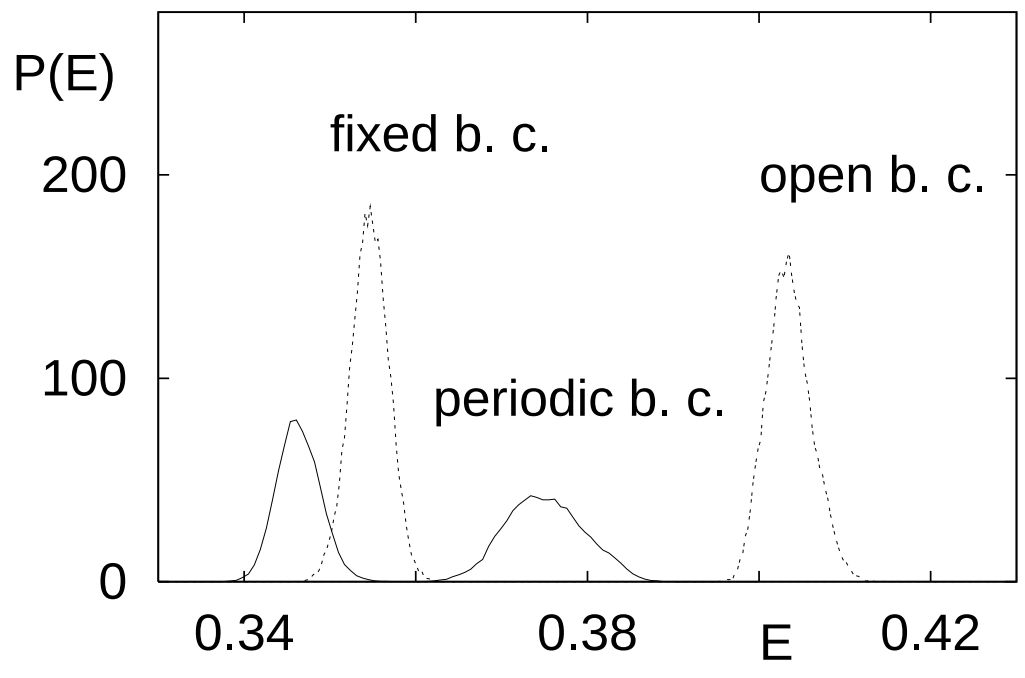

Figure 2. Probability distributions $P(E)$.

To study effects of inhomogeneities we introduce shells (3-dimensional subsets) of the lattice and measure the indicated observables on each of them separately. Our numbering of the shells is $s = 1, \ldots, L/2$ where $L$ is the lattice extension and $s = 1$ corresponds to the outmost shell. A shell then consists of the points where one coordinate is put to $L - s$ or to $s - 1$.

Figure 3 shows the results obtained for each of the shells, separately, in the same simulations as in Figure 2. It is obvious that for open and fixed boundary conditions the values of the observables change dramatically from shell to shell. The directions of the shifts for open and fixed boundary conditions are in accordance with the lattice getting hotter and colder, respectively, for outer shells. In view of such large effects one can hardly consider results about the presence or absence of a gap to be reliable with these types of boundary conditions unless one goes to much larger lattices where the impact of boundaries decreases.

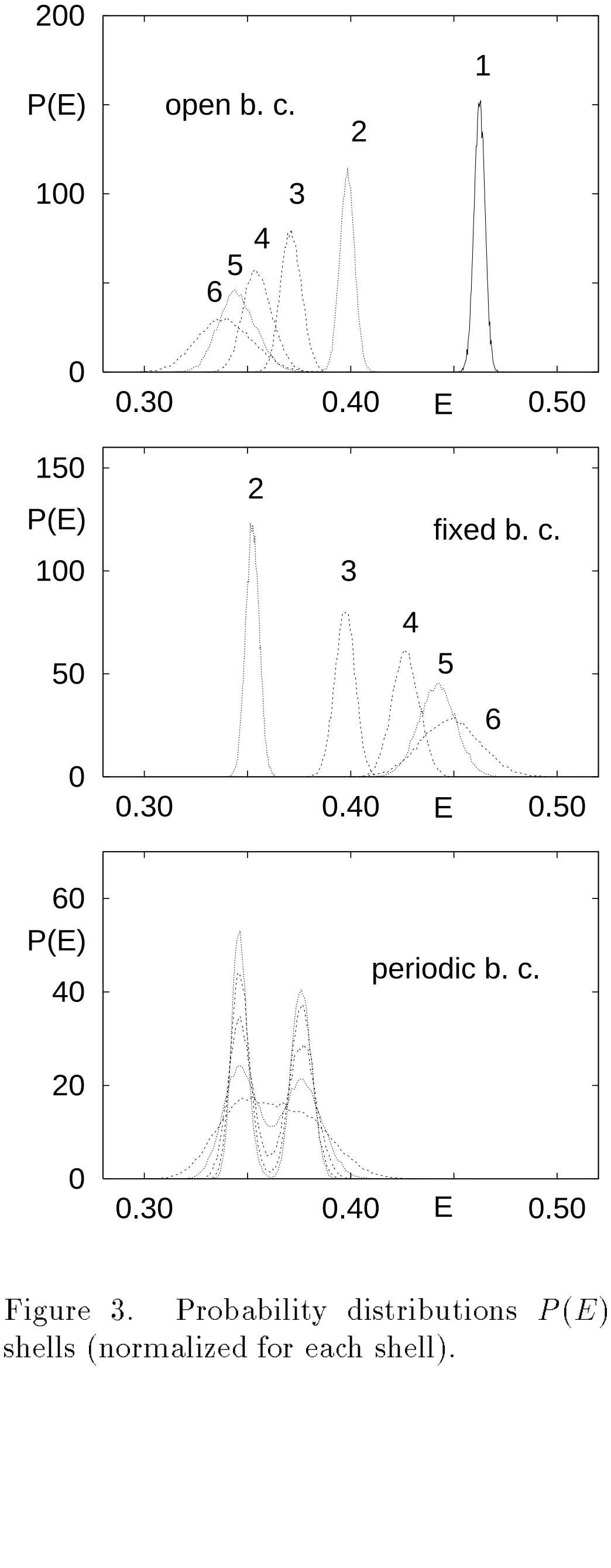

Figure 3. Probability distributions $P(E)$ for shells (normalized for each shell).

## 4. OCCURRENCE OF ENERGY GAP

For periodic boundary conditions an energy gap indicative of a first order transition is seen already on moderately large lattices. While a gap is not yet seen on the $4^4$ lattice, it is there for $8^4$, and for $16^4$ the two peaks of the energy distribution become well separated (in which case our new algorithm [6] makes simulations still possible).

For open and fixed boundary conditions, a gap is not yet seen on a $16^4$ lattice, as is obvious from Figure 2. The widths of the peaks nevertheless remain relatively small which suggests that not simple smearing but a more complicated mechanism is involved. The investigation of larger lattices gets extremely cumbersome. Thus to go to lattice sizes where the observed inhomogenities are no longer important appears hardly feasible.

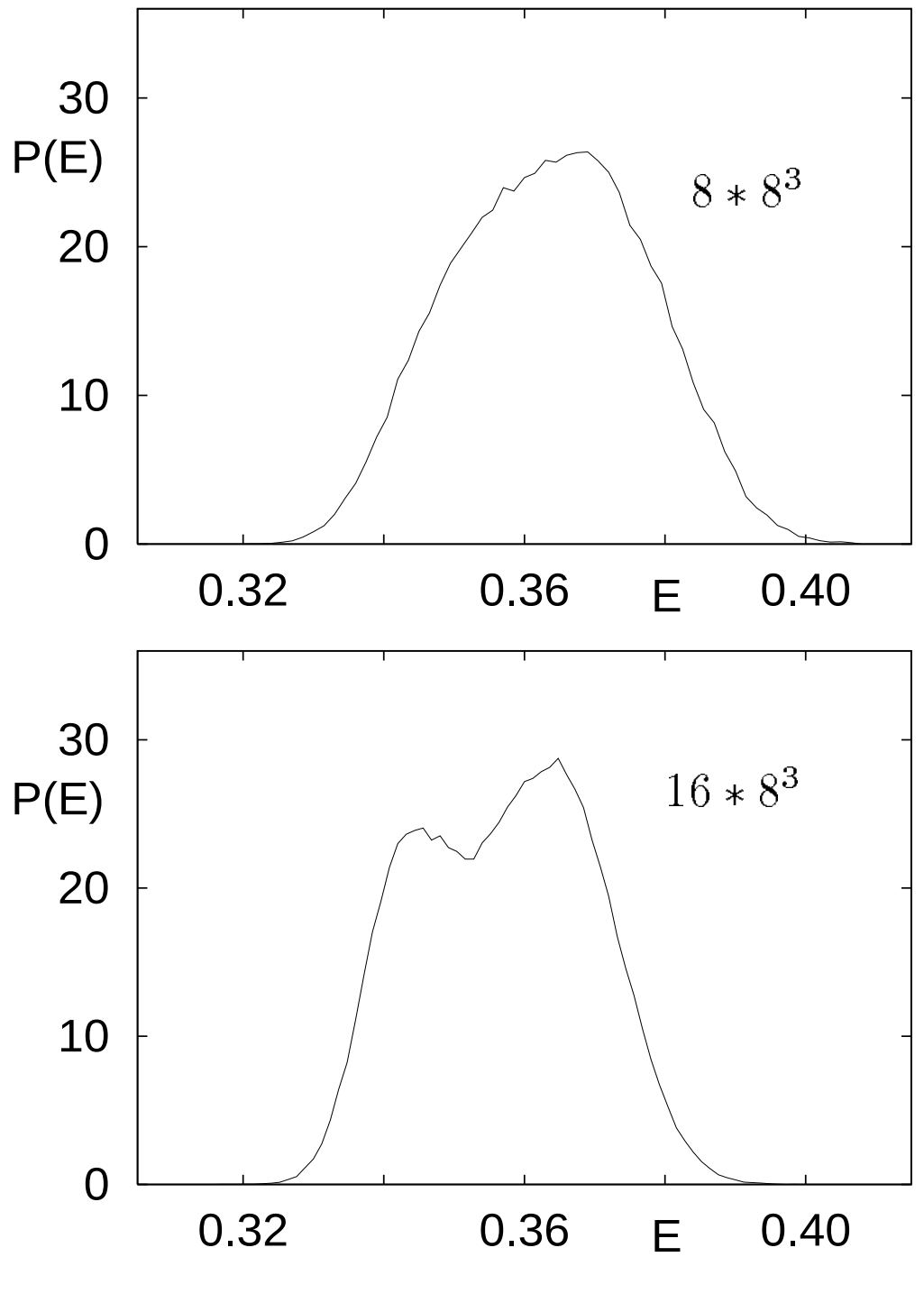

Figure 4. Probability distribution $P(E)$ with fixed boundary conditions in one direction.

Therefore we have looked for other, more manageable settings where to investigate whether the disappearance of the gap is absolute or depends on the size and/or other parameters of the system. In the following we study two cases where such an investigation has been possible.

In the first example we observe that by using fixed boundary conditions only in 0–direction (and periodic ones otherwise) the gap disappears on the $8^4$ lattice. Figure 4 shows our results on a $L_0 \times 8^3$ lattice. While for $L_0 = 8$ there is no gap, for $L_0 = 16$ a gap is seen to occur. Thus in this simplified case the reappearance of the gap for larger lattice size is demonstrated.

In the second example we use fixed boundary conditions in all directions and supplement the Wilson action by a monopole term [7],

$$S = \beta \sum_{\mu>\nu,x} (1 - \cos \Theta_{\mu\nu,x}) + \lambda \sum_{\rho,x} |M_{\rho,x}| \,,$$

where $M_{\rho,x} = \epsilon_{\rho\sigma\mu\nu}(\bar{\Theta}_{\mu\nu,x+\sigma} - \bar{\Theta}_{\mu\nu,x})/4\pi$ and the physical flux $\bar{\Theta}_{\mu\nu,x} \in [-\pi, \pi)$ is given by $\Theta_{\mu\nu,x} = \bar{\Theta}_{\mu\nu,x} + 2\pi n_{\mu\nu,x}$ [9]. From Figure 5 it is seen on a $16^4$ lattice that for sufficiently large monopole density, i.e. negative $\lambda$, a gap emerges. The corresponding results for open boundary conditions are very similar.

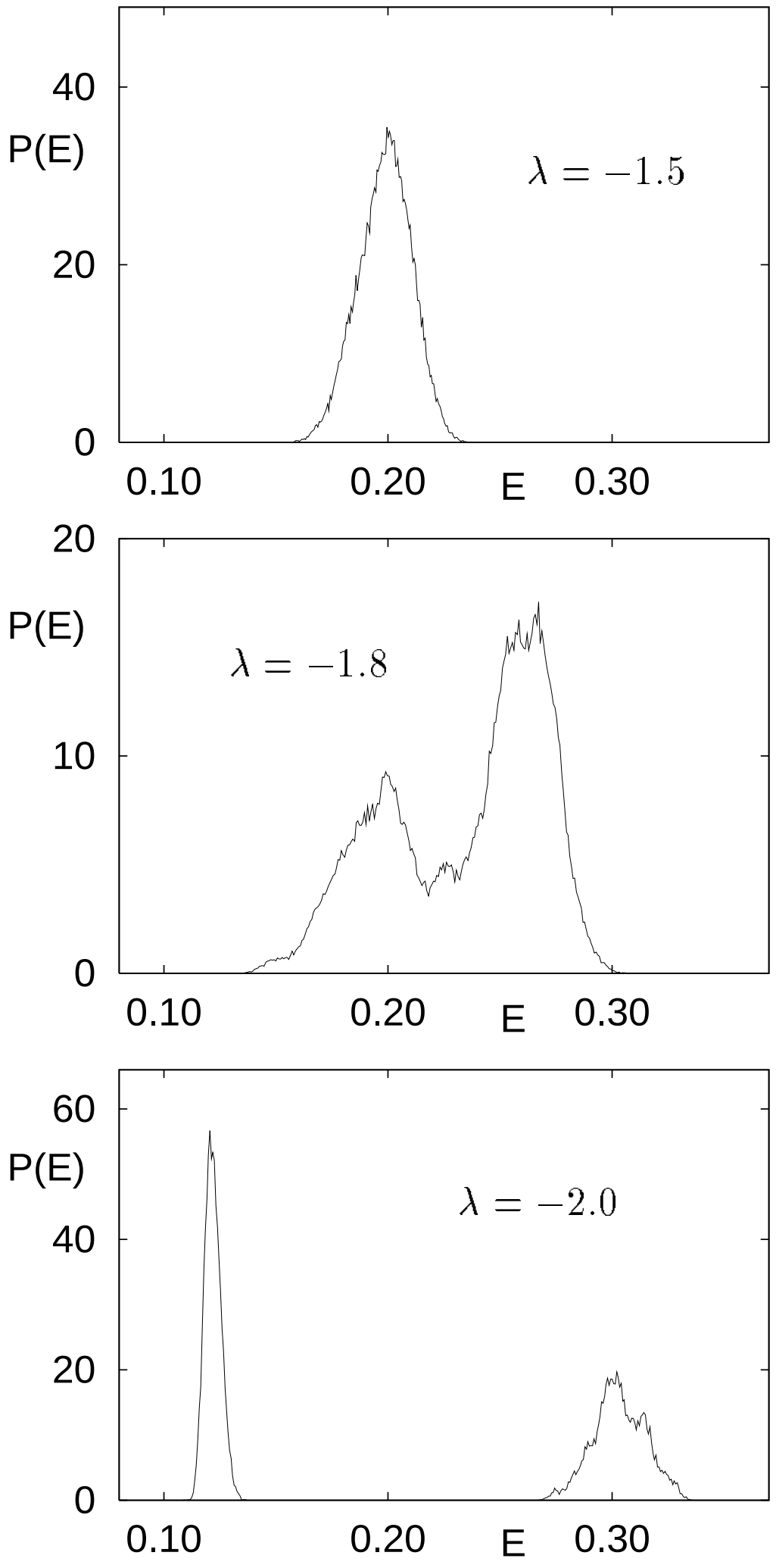

Figure 5. Probability $P(E)$ for fixed boundary conditions and negative $\lambda$.